\documentclass[10pt]{article}
\usepackage{amsfonts, latexsym, amsmath}

\newcommand{\Exp}{{\rm Exp}}

\newcommand{\IR}{\mathbb{R}}
\newcommand{\IC}{\mathbb{C}}

\begin{document}
\begin{center}
{\Large The damped harmonic oscillator in  deformation quantization}
\end{center}
\centerline{
Giuseppe Dito$^{a,}$\footnote{E-mail address: giuseppe.dito@u-bourgogne.fr}
and Francisco J. Turrubiates$^{a,b,}$
\footnote{E-mail addresses: fturrub@u-bourgogne.fr, fturrub@fis.cinvestav.mx}
}

\centerline{\it $^a$Institut de Math\'ematiques de Bourgogne}
\centerline{\it Universit\'e de Bourgogne} \centerline{\it BP
47870, F-21078 Dijon Cedex, France}
\medskip
\centerline{\it $^b$Departamento de F\'{\i}sica} \centerline{\it
Centro de Investigaci\'on y de Estudios Avanzados del IPN}
\centerline{\it Apdo. Postal 14-740, 07000, M\'exico D.F.,
M\'exico}
\vspace{.5cm}
\centerline{October 2005}
\vspace{.5cm}


\vspace{1cm}

\begin{abstract}
We propose  a new approach to the  quantization of the damped harmonic oscillator in the 
framework of deformation quantization. The quantization is performed in the
Schr\"odinger picture by a star-product induced by a modified 
``Poisson bracket''. 
We determine the eigenstates in the damped regime
and compute the transition probability between states of the undamped harmonic 
oscillator after the system was submitted to dissipation.

\end{abstract}
\section{Introduction}

Many attempts to study dissipative systems from the quantum point of view
have been made for many years. Nevertheless, the problem in general is far 
from having a satisfactory solution. 

Quantum Mechanics mainly deals with microscopic systems and since dissipation is a macroscopic
concept some authors pointed out that the quantization problem is not even relevant 
because the nature of interactions or forces involved in such processes are not fundamental \cite{Senit}.
However there are problems where the dissipation at this level has an important role as in
the quantum analysis of the radiation field in a microwave cavity, in 
quantum optics and  also for {\it deep
inelastic processes} in heavy ion scattering, to mention some of them \cite{Senit,Hass}.

The damped harmonic oscillator (DHO) is one of the simplest cases of this type treated
in the literature (for a review  on the quantization problem see 
\cite{Dek} and the references therein).
Most of the attempts have been using the canonical quantization formalism, 
dealing either with Hamiltonians dependent or independent of time, or by duplicating the 
number of degrees of freedom to obtain a Hamiltonian system that can be quantized 
by standard methods. We believe that the quantization of the DHO from first principles 
should bring new insights and ideas for a better understanding of
the quantum aspects of dissipative phenomena. 

In this work we propose a quantization of the DHO from the point of view
of deformation quantization.
Our approach consists to include all of the dissipative effects into the algebra 
of observables. This is achieved by defining a star-product that takes into account
the dissipation character of the DHO. We follow an idea of Enz \cite{Enz} to add
a symmetric part to the Poisson bracket (mathematically, a Hochschild coboundary)
and work with the star-product $\star_\gamma$ induced by this modified bracket.

Chru\'sci\'nski \cite{Chru, Chru2} has also considered the quantization problem
for dissipative systems within the deformation quantization framework (see also \cite{Taras}). 
His treatment differs from ours in the sense that he deals 
with the Moyal product and we rather use a mathematically equivalent star-product. 
In the limit when the damping factor goes to $0$, we recover the 
Moyal product.

Our paper is organized as follows. In Section~2 we review  fundamental notions of
deformation quantization and give some details on the quantization of the 
harmonic oscillator in both Schr\"odinger and Heisenberg pictures. 
We introduce in Section~3 a new bracket between functions by adding a 
``dissipative'' symmetric part to the canonical Poisson bracket. 
The modified bracket induces a $\star$-product that provides a quantization 
of the damped harmonic oscillator and allows obtain the corresponding quantum states. 
In Section~4, we compute some relevant transition probablities and 
observe that the dissipation couples a given eigenstate to lower energy states.

\section{Deformation quantization and the harmonic oscillator}

In this section we briefly introduce the basic notions of the deformation quantization 
formalism. We review the quantization of the harmonic oscillator in this setting
by putting the emphasis on the Moyal-Schr\"odinger picture that will be used later on
for the damped case. 

\subsection{Notions on deformation quantization}

Deformation quantization was introduced in \cite{Bayen} as an alternative approach 
to quantization. In this formalism, quantization is interpreted as a deformation
of the algebra of smooth functions on the classical phase space. We refer the
reader to the original papers \cite{Bayen} for details 
and the review \cite{DitSt} for more recent developments.

The  phase space of a classical system is given by a manifold $M$
endowed with a Poisson bracket $P$ on the algebra of complex-valued smooth 
functions $C^{\infty}(M)$. We shall denote by $C^{\infty}(M)[[\hbar]]$ the space 
of formal series in $\hbar$ with coefficients
in $C^{\infty}(M)$. An element $f$ in $C^{\infty}(M)[[\hbar]]$ can be written as:
\begin{equation}\label{zero}
f=\sum_{k=0}^{\infty} \hbar^k f_k
\end{equation}
where the $f_k$'s are in $C^{\infty}(M)$. Rather than the Planck constant $\hbar$, 
we shall use for physical reasons $i\hbar/2$  $(i^2=-1)$ as deformation parameter.
\vskip5mm
\noindent {\bf Definition}. {\it A deformation quantization of a classical 
system defined by $(M,P)$ is an associative algebra $(C^{\infty}(M)[[\hbar]],\star)$ 
over $\IC[[\hbar]]$, where the associative product $\star$, the so-called star-product,}
\begin{equation}\label{one}
f \star g = \sum_{k=0}^{\infty}(i\hbar/2)^k C_k(f,g), \quad f,g \in C^{\infty}(M),
\end{equation}
{\it satisfies:}
\vskip1mm
\noindent a) {\it $C_0(f,g) =fg$, and the $C_k$'s, $k \geq 1$,
are bidifferential operators on $C^{\infty}(M)$,
\vskip1mm
\noindent b) the constant function $1$ is the unit for $\star$, i.e., 
$1\star f=f\star 1=f$, for any smooth function $f$,
\vskip1mm
\noindent c) $C_1(f,g)-C_1(g,f)= 2 P(f,g)$, so that the star-bracket
 $[f,g]_\star \equiv  (f \star g - g \star f)/i\hbar$ is a deformation of the 
classical Poisson bracket $P$.
}\vskip1mm
It is a famous result of Kontsevich \cite{Kont} that any Poisson manifold $(M,P)$
admits a  deformation quantization.

The algebra of classical observables $\cal{O}$ is the set of real-valued functions 
on the phase space. Given a classical Hamiltonian $H\in \cal{O}$, 
the quantum evolution of an observable $f$ is still governed by the classical 
Hamiltonian, but one  replaces the Poisson bracket by the star-bracket in order
to get the Heisenberg equation of motion:
\begin{equation}\label{zerobis}
\frac{df}{dt}= [f,H]_\star = \frac{f \star H - H \star f}{i\hbar}=P(f,H) 
+\ {\rm corrections\ in}\ \hbar.
\end{equation}
As $\hbar\rightarrow0$, one recovers in some sense the classical equation of motion. 
The heuristic solution to Eq.~(\ref{zerobis}) is given by 
$f_t=\Exp_{\star}\big(\frac{-tH}{i\hbar}\big)\star f 
\star \Exp_{\star}\big(\frac{tH}{i\hbar}\big)$, where
\begin{equation}\label{five}
\Exp_{\star}\big(\frac{tH}{i\hbar}\big) \equiv \sum_{n=0}^\infty 
\frac{1}{n!}\big(\frac{t}{i\hbar}\big)^n H^{\star n},
\end{equation}
with $H^{\star n} = H \star\cdots \star H$ ($n$  factors). 
The star-exponential~(\ref{five}) is in general a quite singular object, 
nevertheless, for many physically relevant examples, the series converges 
as a distribution (see \cite{Bayen}). An important feature of 
the star-exponential is that its Fourier transform (when it exists in a distribution sense): 
$\int e^{-itE/\hbar} d\nu(E)$ allows to identify the spectrum of $H$ as the support
of the measure $\nu$. More precisely, suppose for simplicity that 
the Hamiltonian  is such that the series in (\ref{five}) defines a periodic distribution
in $t$, then the measure $\nu$ is atomic and the Fourier transform reduces to a Fourier series
(i.e. discrete spectrum):
\begin{equation}\label{fivefourier}
\Exp_{\star}\big(\frac{tH}{i\hbar}\big)= \sum_{n=0}^\infty 
e^{\frac{t E_n}{i\hbar}} \pi_n .
\end{equation}
The $E_n$'s are the eigenvalues of $H$ and the functions $\pi_n$ 
represent the orthonormal eigenstates in the sense that:
\begin{equation}\label{eigen}
H\star \pi_n = E_n \pi_n, \quad \pi_m\star \pi_n = \delta_{mn}\pi_n.  
\end{equation}
Strictly speaking this makes sense if one is using a convergent form of the product $\star$.
One may think of $\pi_n$ as the function sent by the quantization rule 
to the projection operator $|n\rangle\langle n|$
in the Heisenberg picture of Quantum Mechanics, $\{|n\rangle\}$ being 
the normalized eigenstates of the Hamiltonian operator $\hat H$ corresponding to $H$.

\subsection{The harmonic oscillator in the Moyal-Heisenberg picture}

Let us consider the Hamiltonian $H(q,p)= \frac{1}{2m} p^2  + \frac{m \omega^2}{2} q^2$
of the harmonic oscillator on the phase space  $\IR^2$, with coordinates $(q,p)$,
endowed with the canonical Poisson bracket
\begin{equation}\label{three}
P(f,g)=
\frac{\partial f}{\partial q}\frac{\partial g}{\partial p} 
-  \frac{\partial f}{\partial p} \frac{\partial g}{\partial q}.
\end{equation}
The Moyal product between two functions $f,g \in C^{\infty}(\IR^2)$ is given by
the formal series:
\begin{equation}\label{four}
f \star g =  \exp\big(\frac{i \hbar}{2} P\big)(f, g) \equiv fg + 
\sum_{k\geq1} \frac{1}{k!}\big(\frac{i \hbar}{2}\big)^k P_k(f,g),
\end{equation}
where $P_k$ is the $k^{\rm th}$-power of $P$, defined by
\begin{equation}\label{kpower}
P_k(f,g)(q,p) = \big( \frac{\partial}{\partial q_1}\frac{\partial}{\partial p_2}
 - \frac{\partial}{\partial p_1} \frac{\partial}{\partial q_2}
\big)^k f(q_1,p_1)g(q_2,p_2)\Big|_{\genfrac{}{}{0pt}{}{q_1=q_2=q}{p_1=p_2=p}}.
\end{equation}

The star-exponential for the harmonic oscillator has been computed in \cite{Bayen}
and the series~(\ref{five}) converges as a distribution to
\begin{equation}\label{six}
\Exp_{\star}\big(\frac{tH}{i\hbar}\big)= \big(\cos(\frac{\omega t}{2}) \big)^{-1} 
\exp{\big( \frac{2H}{i\hbar \omega}
\tan(\frac{\omega t}{2}) \big)}.
\end{equation}
The Fourier expansion of the right-hand side of (\ref{six}) is given by \cite{Bayen}:
\begin{equation}\label{seven}
\Exp_{\star}\big(\frac{tH}{i\hbar}\big) = \sum_{n=0}^\infty 
\exp{\big( -i(n+1/2)\omega t \big)} \pi_n,
\end{equation}
where 
\begin{equation}\label{eight}
\pi_n(q,p)=2 \exp{\big( \frac{-2 H(q,p)}{\hbar\omega}\big)}
(-1)^n L_n \big( \frac{4 H(q,p)}{\hbar\omega}\big),
\end{equation}
and where $L_n$ is the Laguerre polynomial of degree $n$. In particular, the Gaussian
function 
\begin{equation}\label{pio}
\pi_0(q,p) = 2 \exp{\big( \frac{-2 H(q,p)}{\hbar\omega}\big)}
\end{equation}
describes the vacuum state and,
as expected, the energy levels of $H$ are $E_n=\hbar \omega (n + 1/2)$. All this is well known
and gives the essence of the deformation quantization of the harmonic oscillator in the Heisenberg
picture. We shall below consider the less-known Schr\"odinger picture in deformation quantization.

\subsection{The harmonic oscillator in the Moyal-Schr\"odinger picture}

In order to write the Schr\"odinger equation:
\begin{equation}\label{schro}
i\hbar \frac{\partial \psi(t)}{\partial t} = H\star \psi(t),
\end{equation}
it is implicitly assumed that we have picked up some representation of the algebra of observables
$(\mathcal{O},\star)$, so that we can `multiply' observables with states. We refer the reader
to the excellent review by Waldmann \cite{Wald} and references therein for a comprehensive
exposition on  states and representations in deformation quantization.

We shall consider quantum states as a class of functions on the phase space. Strictly speaking
this make sense only if we have a convergent (integral) form of the Moyal product $\star$ between
a certain class of functions and $\hbar$ should no longer be a formal parameter. 
Indeed, solutions of the eigenvalue equation
$ H\star \psi=E \psi$ are not formal series in $\hbar$, and it does not make sense to
multiply such functions with the product $\star$ as  formal series. Though we do not
want to present a mathematical analysis of the convergence of the product $\star$ here, let us
only mention that a convergent form does exist and gives a meaning to the product in such cases.

For convenience, let us introduce the (dimensionless) annihilation and creation functions:
\begin{equation}\label{ten}
a(q,p)=\frac{p}{\sqrt{2m \hbar \omega}} - i \sqrt{\frac{m \omega}{2 \hbar}} q,\quad 
\bar a(q,p) = \frac{p}{\sqrt{2m \hbar \omega}} + i \sqrt{\frac{m \omega}{2 \hbar}} q.
\end{equation}
It is easily checked that we have:
\begin{equation}\label{tenbis}
H = \hbar\omega \bar a a= \hbar\omega (\bar a\star a + 1/2),\quad [a,H]_\star= -i\omega a,\quad 
\quad [\bar a,H]_\star= i\omega \bar a,\quad \quad [a,\bar a]_\star= \frac{1}{i\hbar}.
\end{equation}

The vacuum $\rho_0$ in the Moyal-Schr\"odinger picture is the same as 
the one provided by the star-exponential (\ref{pio}), i.e., $\rho_0=\pi_0$, and
in view of the relations (\ref{tenbis}), it is clear that 
$\tilde\rho_n\equiv \bar a \star\cdots \star \bar a \star \rho_0$ (there are $n\ \bar a$'s) 
solve the eigenvalue problem: $H\star \tilde\rho_n =E_n \tilde\rho_n$ with $E_n=  \hbar \omega (n+1/2)$.
A simple computation using the Moyal product gives $\tilde\rho_n = 2^n \bar a^n \rho_0$.

In order to get a Hilbert space of states, we shall normalize the $\tilde\rho_n$'s with respect
to the scalar product:
\begin{equation}\label{scalar}
\langle f | g \rangle^{}_{L^2} = \int_{\IR^2} \bar f(q,p) g(q,p) d\mu(q,p),
\end{equation}
where $ d\mu(q,p)= \frac{dqdp}{2\pi \hbar}$ is the Liouville measure. A  computation
using Gaussian integrals shows that:
\begin{equation}\label{states}
\rho_n = \frac{1}{\sqrt{n!}}\ \bar a^n \rho_0,\quad n\geq0,
\end{equation}
is an orthonormal set for the scalar product~(\ref{scalar}). 

The vector space over $\IC$ generated by $\{\rho_n\}_{n\geq0}$ is 
$S_M\equiv\{h(\bar a)\pi_0 | h \ {\rm polynomial} \}$ and by taking the closure of 
$S_M$ with respect to the scalar product~(\ref{scalar}), one gets the Hilbert
space ${\cal H}_M=\{ f(\bar a) \pi_0 | f \ {\rm entire\ function\ such\ that} 
\int |f(\bar a(q,p))|^2 e^{\frac{-4H}{\hbar \omega}}d\mu(q,p)< \infty \}$. Since
$H=\hbar\omega \bar a a$, it is clear that ${\cal H}_M$ is isometric to the Bargmann-Fock space 
${\cal H}_{BF}$ of antiholomorphic functions $f$ on $\IC$ such that:
\begin{equation}\label{bfscalar}
\langle f| f\rangle^{}_{BF} \equiv\frac{2}{\pi}\int_\IC  |f(\bar \xi)|^2 e^{-4 \bar \xi \xi} d\xi d\bar \xi  <  \infty.
\end{equation}
Notice that $S_M$ can be seen as a dense invariant {\it domain} for $H$ in the sense that $H\star S_M \subset S_M$.
Any state $\psi$ can be expanded on the basis~(\ref{states}): $\psi(q,p)=\sum_{n\geq0} \alpha_n \rho_n(q,p)$, 
where $\alpha_n\in\IC$ and $\sum_{n\geq0} |\alpha_n|^2=1$ in full analogy with standard Quantum Mechanics.


\section{The damped harmonic oscillator}

We shall introduce a new star-product $\star_\gamma$ that will provide a quantization 
in the Schr\"odinger picture of the damped harmonic oscillator whose dynamics is governed by:
\begin{equation}\label{setentaq}
\ddot{q} + 2 \gamma \dot{q} + \omega^2 q = 0,
\end{equation}
where  $\gamma>0$ denotes the damping factor and $\omega$ is the frequency.

Our philosophy  is to incorporate all of the dissipative effects in the algebra 
of observables $\mathcal{O}_\gamma$ by an appropriate deformation of the 
Moyal product, with $\gamma$ as a second deformation parameter. 
We would like not to deform the measurable quantities and put everything
related with dissipation into the star-product $\star_\gamma$.
For example, the function $H= \frac{1}{2m} p^2  + \frac{m \omega^2}{2}  q^2$ still represents 
the mechanical energy of the particle submitted to dissipation. At the quantum level,
we would like to keep $H$ as the generator of some $\gamma$-deformed dynamics 
so that (\ref{setentaq}) will still be deduced from the usual undamped Hamiltonian
in a sense that we will now make precise.

We mention that a singular Hamiltonian for the underdamped case has been derived in \cite{PC}.
However the Hamiltonian considered there is only locally defined on the phase space. Consequently,
the solutions of the induced system of equations are defined for small times only and cannot be extended globally in time.

\subsection{The modified bracket}

Consider $\IR^2$ as the phase space with coordinates $(q,p)$. Eq~(\ref{setentaq}) is equivalent
to the system of equations:
\begin{equation}\label{dhotwo}
\dot{q}  = \frac{p}{m} ,\qquad
\dot{p} =  - m \omega^2 q - 2 \gamma p.
\end{equation}

We shall rewrite these equations in a form similar to the Hamilton equations of motion
by using a modified bracket $M$ obtained by adding a symmetric part to the canonical Poisson
bracket $P$. This idea to consider a modified bracket for the quantization
of dissipative systems within the Heisenberg picture is due to  Enz \cite{Enz}. 

It is easy to check that the only bracket $M$ that is a derivation in each of its argument and
such that Eq.~(\ref{dhotwo}) can be cast into a form mimicking the Hamilton equations of motion:
\begin{eqnarray}\label{twoa}
\dot{q} & = & M(q,H)= \frac{p}{m},  \nonumber \\
\dot{p} & = & M(p,H)= - m \omega^2 q - 2 \gamma p,
\end{eqnarray}
is given by
\begin{equation}\label{dhoone}
M(f,g)= \frac{\partial f}{\partial q} \frac{\partial g}{\partial p} - 
\frac{\partial f}{\partial p} \frac{\partial g}{\partial q} -
2 \gamma m \frac{\partial f}{\partial p} \frac{\partial g}{\partial p}.
\end{equation}

Here is another motivation for this modified bracket and justification for the form of Eq.~(\ref{twoa}). 
Let us consider a solution $t\mapsto (x(t),y(t))$ of Eq.~(\ref{dhotwo}) satisfying 
the initial conditions $x(0)=q$ and  $y(0)=p$.
The evolution of any smooth function $f\in C^\infty(\IR^2)$ is given by the
path $t\mapsto f_t$ in $C^\infty(\IR^2)$ defined by $f_t(q,p)\equiv f(x(t),y(t))$.
The evolution equation of $f_t$ is deduced by taking its time derivative and using Eq.~(\ref{dhotwo}):
\begin{eqnarray*}
\frac{d}{d t} f(x(t),y(t)) & = & \frac{\partial f}{\partial q}(x(t),y(t))\ \dot x(t) 
+ \frac{\partial f}{\partial p}(x(t),y(t))\  \dot y(t)   \\
& = &  \frac{\partial f}{\partial q}(x(t),y(t))\ \frac{y(t)}{m}
- \frac{\partial f}{\partial p}(x(t),y(t))\ (m\omega^2 x(t) + 2 \gamma y(t)), \\
 & = & P(f,H)(x(t),y(t))  
- 2 \gamma m \frac{\partial f}{\partial p}(x(t),y(t))\ \frac{\partial H}{\partial p}(x(t),y(t))\\
& = & M(f,H)(x(t),y(t)).
\end{eqnarray*}
Thus the introduction of the modified bracket $M$ appears to be quite  natural.
In particular, the time evolution of the mechanical energy $H$ is
\begin{equation}
\dot{H} = M(H,H) = -\frac{2\gamma}{m} p(t)^2 \leq 0,
\end{equation}
and exhibits energy loss due to dissipation.

\subsection{The star-product $\star_\gamma$}

{}From the mathematical and deformation theory points of view, the bracket $M$
has the following important property: it is a constant coefficient bidifferential
operator of order $1$ in each of its argument. This implies that $M$ is
a constant coefficient Hochschild 2-cocycle (see e.g. \cite{DitSt} for details on the Hochschild cohomology).
As an immediate consequence of the previous fact,
the following formula:
\begin{equation}\label{dhoseven}
f \star_{\gamma} g = \exp (\frac{i\hbar}{2} M) (f,g)
\end{equation}
defines a formal associative product \cite{Bayen}, i.e., a star-product. 
The formula~(\ref{dhoseven}) should be interpreted in the usual way:
$f\star_{\gamma} g=fg + \sum_{k\geq1} \frac{1}{k!} \big(\frac{i\hbar}{2}\big)^k M_k(f,g)$, where  $M_k$
is the $k^{\rm th}$-power of $M$ defined in a similar fashion as in (\ref{kpower}).

The star-product appears as a $\gamma$-deformation
of the Moyal product $\star_M$ in the sense that in the limit $\gamma \rightarrow 0^+$ we indeed
recover the Moyal product.
Actually the relation between the product $\star_{\gamma}$ and the Moyal product can be made explicit.
First notice that the bracket $M$ can also be written as:
\begin{equation}\label{dhofour}
M(f,g)= P(f,g) + 2 m \gamma  \delta \theta (f,g),
\end{equation}
where $\theta (f)= \frac{1}{2} \frac{\partial ^2 f}{\partial p^2}$, and $ \delta \theta$
is the  Hochschild differential of $\theta$ defined by 
$\delta \theta(f,g)= f \theta(g) - \theta(fg) + \theta(f)g$.
It implies that the product $\star_{\gamma}$ is mathematically equivalent \cite{Bayen} to the Moyal
product, i.e., there exists a formal series of differential operators 
$T = Id + \sum_{r\geq1} (\frac{i\hbar}{2})^r T_r$ so that $T(f \star_M g) = T(f) \star_\gamma T(g)$.
Here the form~(\ref{dhofour}) of $M$ gives us that:
\begin{eqnarray}\label{dhoeight}
T & = & \exp (- i{\hbar} m \gamma  \theta)  = \exp \big(\frac{-i\hbar m \gamma}{2} 
\frac{\partial^2}{\partial p^2}\big).
\end{eqnarray}

We shall use the product  $\star_{\gamma}$ to quantize the damped harmonic oscillator. Notice
that $\star_{\gamma}$ is not a Hermitian product \cite{Wald}: 
$\overline{f \star_{\gamma} g}\neq \bar{g} \star_{\gamma} \bar{f}$,
where the bar denotes complex conjugation. In fact the operator $T$ destroys the Hermitian nature
of the Moyal product as $\overline{T(f)}\neq T(\bar{f})$ and this is precisely what is needed  for
dissipative phenomena.

\subsection{The quantum states}

The Schr\"odinger equation in our context reads:
\begin{equation}\label{schrobis}
i\hbar \frac{\partial \psi(t)}{\partial t} = H\star_\gamma \psi(t).
\end{equation}
The remark after (\ref{schro}) also applies to the product $\star_\gamma$. 

Heuristically, the solution of the Schr\"odinger equation is given by:
\begin{equation}\label{starevol}
\psi(t) = U(t)\star_\gamma \psi(0),
\end{equation}
where $U(t)=  \Exp_{\star_\gamma}\big( \frac{t H}{i\hbar}\big)$ is the star-exponential of $H$ for
the product $\star_\gamma$. The function $U(t)$ can be in principle determined by solving the differential
equation:
$$
i\hbar \frac{\partial U(t)}{\partial t} = H\star_\gamma U(t),\quad U(0)=1,
$$
but it is easier to deduced it from the star-exponential~(\ref{six}) 
of $H$ for the Moyal product and the use of the equivalence operator~(\ref{dhoeight}). 
A long but straightforward computation gives:
\begin{eqnarray}\label{dhoten}
\Exp_{\star_\gamma}\big( \frac{t H}{i\hbar}\big)
&=& \frac{\exp(\gamma t/2)}{\cos(\omega t/2)  (1 + \frac{2 \gamma}{\omega} \tan (\omega t/2))^{1/2}} 
\times \\
& & \times \exp \bigg(\frac{- i}{\hbar\omega}\tan(\omega t/2)
\big(m \omega^2 q^2 + \frac{p^2}{m(1 + \frac{2 \gamma}{\omega} \tan (\omega t/2))}\big)\bigg).\nonumber
\end{eqnarray}
According to the general principle of deformation quantization \cite{Bayen}, the spectrum and the
states for the Heisenberg picture are obtained by performing a Fourier decomposition of
the star-exponential. Hence  the periodic term in $t$ in Eq.~(\ref{dhoten}), can be expanded as
a Fourier series:
\begin{equation}\label{eleven}
\Exp_{\star_\gamma}\big( \frac{t H}{i\hbar}\big)(q,p) = \exp (\gamma t/2) \sum_{n\geq 0} 
\exp \big(\frac{t \lambda_n}{i \hbar}\big) \pi^\gamma_n (q,p).
\end{equation}
Again using the equivalence operator~(\ref{dhoeight}), one finds that $\lambda_n=\hbar\omega(n+1/2)$
and thus the energy levels are given by $E^\gamma_n = \hbar \omega(n+ 1/2 + i \gamma/ 2\omega)$.
Notice that the spectrum is now complex as a consequence of the fact that we are dealing with a nonhermitian product
or, equivalently, an eventual operator representation of the algebra of observables for the product 
$\star_\gamma$ would represent $H$ as a non-selfadjoint operator.

The eigenstates $ \pi^\gamma_n (q,p)$ in the Heisenberg picture can also be computed with the
help of (\ref{dhoeight}), but we just need to do it for the vacuum $\pi^\gamma_0$, and it is given by:
\begin{equation}\label{dhovac}
\pi^\gamma_0 (q,p)=  \frac{2}{(1 - i 2 \gamma/\omega)^{1/2}} 
\exp\big(\frac{-1}{\hbar \omega}(m \omega^2 q^2 + \frac{p^2}{m( 1  - i 2 \gamma/\omega)})\big).
\end{equation}
(All over the paper, the branch cut of the square root is along the negative real axis.)
As $\gamma\rightarrow 0^+$, we recover the vacuum~(\ref{pio}) for the Moyal case.
It turns out that $\pi^\gamma_0$ is normalized for the $L^2$-scalar product~(\ref{scalar}):  
$\int |\pi^\gamma_0|^2 d\mu(q,p)=1$ and it is annihilated by $a(q,p)$, i.e., $a\star_\gamma \pi^\gamma_0 =0$.
Moreover we have:
\begin{equation}\label{dhotenbis}
H = \hbar\omega (\bar a\star_\gamma a + 1/2 + i\gamma/2\omega),\quad [a,H]_{\star_\gamma}= -i\omega a,\quad 
\quad [\bar a,H]_{\star_\gamma} = i\omega \bar a,\quad \quad [a,\bar a]_{\star_\gamma}= \frac{1}{i\hbar}.
\end{equation}
As in the Moyal case, it follows from the preceding relations that the fonctions:
$$
\tilde\rho^\gamma_n = \bar a\star_\gamma\cdots  \star_\gamma \bar a\star_\gamma \rho^\gamma_0,\quad
(\rho^\gamma_0\equiv \pi^\gamma_0)
$$ 
solve the $\star_\gamma$-eigenvalue problem for $H$, i.e., $H\star_\gamma \tilde\rho^\gamma_n
= \hbar\omega (n+ 1/2 + i \gamma/ 2\omega) \tilde\rho^\gamma_n$. 

The $\tilde\rho^\gamma_n$'s can be explicitly computed by taking the $n^{\rm th}$ derivative
in the variable $s$ of the generating function
$G(s,q,p) = \sum_{n\geq 0} \frac{s^n}{n!} \tilde\rho^\gamma_n(q,p)$. $G(s,q,p)$ satisfies the differential equation:
\begin{eqnarray*}
\frac{\partial G}{\partial s}&=& \bar a \star_\gamma G,\\
 &=& \bar a G - \frac{i}{2\sqrt{2}}\sqrt{\frac{\hbar}{m \omega}} \frac{\partial G}{\partial q}
- \frac{1}{2\sqrt{2}}\sqrt{m\hbar\omega}\big(1 + \frac{i2\gamma}{\omega}\big)\frac{\partial G}{\partial p},
\end{eqnarray*}
with the initial condition $G(0,q,p) = \rho^\gamma_0(q,p)$, the solution of which is found
by elementary means to be:
\begin{equation}\label{gene}
G(s,q,p) = \exp\big( 2 s \bar a(q, \frac{p}{1 - i 2 \gamma/\omega}) 
-i s^2 \frac{\gamma/\omega}{1 - 2 i \gamma/\omega} \big)\rho^\gamma_0(q,p). 
\end{equation}
The exponential factor in~(\ref{gene}), after a suitable scaling of the variable $s$, 
is the generating function of the Hermite polynomials $H_n$: 
$\exp(2 s z - s^2) =\sum_{n\geq0} \frac{s^n}{n!} H_n(z)$, and one easily find that:
\begin{equation}\label{orthogo}
\tilde\rho^\gamma_n(q,p) = \big(\frac{i\gamma/\omega}{1- i 2\gamma/\omega} \big)^{n/2} 
H_n\big((\frac{1- i 2\gamma/\omega}{i\gamma/\omega})^{1/2}\ \bar a(q, \frac{p}{1 - i 2 \gamma/\omega})\big)
\rho^\gamma_0(q,p).
\end{equation}
The  term  of degree $n$ in $H_n(z)$ is $2^n z^n$, and in the limit $\gamma\rightarrow 0^+$, we indeed recover
$\tilde\rho_n = 2^n \bar a^n \rho_0$, the unnormalized states obtained before for the Moyal product.
The normalized eigenstates for the scalar product~(\ref{scalar}) are given by:
\begin{equation}\label{orthonor}
\rho^\gamma_n(q,p) = \frac{\Gamma^{n/2}}{\sqrt{n!}} 
H_n\big(\frac{1}{\Gamma^{1/2}}\ \bar a(q, \frac{p}{1 - i 2 \gamma/\omega})\big)
\rho^\gamma_0(q,p),
\end{equation}
where $\Gamma = \frac{i\gamma/\omega}{1- i 2\gamma/\omega}$, and thus we have:
\begin{equation}\label{resume}
H\star_\gamma \rho^\gamma_n = \hbar\omega (n+ 1/2 + i \gamma/ 2\omega) \rho^\gamma_n,\qquad
\langle \rho^\gamma_j | \rho^\gamma_{j'} \rangle^{}_{L^2} = \delta_{jj'}.
\end{equation}

The Hilbert space ${\cal H}_{\gamma}$ for the damped harmonic oscillator is obtained
in a natural way by taking the closure with respect to the $L^2$-scalar product
of the vector space generated by $\{\rho^\gamma_n\}$. Let us introduce the complex-valued function
$b(q,p)$ whose complex conjugate is defined by $\bar b(q,p)= \bar a(q, \frac{p}{1 - i 2 \gamma/\omega})$.
Hence ${\cal H}_{\gamma}$ consists
of  functions $f(\bar b(q,p))\rho^\gamma_0(q,p)$, where $f$ is a holomorphic function in the variable $\bar b$
so that $\langle f(\bar b)\rho^\gamma_0 | f(\bar b)\rho^\gamma_0 \rangle^{}_{L^2}<\infty$.
Notice that $H$ has a complex spectrum so it does not lead a self-adjoint operator on ${\cal H}_{\gamma}$,
one can check that $\langle H\star\phi | \psi \rangle^{}_{L^2}\neq \langle \phi | H\star\psi \rangle^{}_{L^2}$
in general.


\section{Measurements}

In the previous section, we have obtained an algebraic description of the dissipative effects 
of the damped harmonic oscillator by  means of the product $\star_\gamma$. 
In order to get a better understanding of the $\star_\gamma$ quantization during dissipation, we shall
consider the following simple toy-model.

Consider the usual one-dimensional harmonic oscillator. At time $t=0$, one switches on the dissipation ($\gamma > 0$) 
and let the system evolves until time $t=\tau>0$, at which time the dissipation is switched off so that
we recover again the undamped case ($\gamma = 0$).

Before $t=0$, we have the usual harmonic oscillator described in terms of the Moyal star product $\star_M$ 
and the eigenstates $\rho_n(q,p)$ given in (\ref{states}). For  $0<t<\tau$, the quantum dynamics is governed
by the product $\star_\gamma$ and the eigenstates $\rho^\gamma_n(q,p)$ appearing in (\ref{orthonor}), and
for $t>\tau$ the dynamics is given again by the Moyal product.

Suppose that the system is known to be in the eigenstate $\rho_n(q,p)$ of the harmonic oscillator for $t<0$ (the phase
is chosen so that it equals 1 for $t=0$).
This state belongs to the Hilbert space ${\cal H}_M$ defined at the end of Sect.~2.3. As the damping factor
$\gamma$ is turned on at $t=0$, the state of the system would then be in the Hilbert space ${\cal H}_\gamma$ of
the DHO. At $t=0$, we use the following prescription
\begin{equation}\label{pres}
\rho_n(q,p)= \frac{1}{\sqrt{n!}} (\bar a(q,p))^n \rho_0(q,p)\rightarrow  
\frac{1}{\sqrt{n!}} (\bar b(q,p))^n \rho^\gamma_0(q,p)
\end{equation}
to express states in ${\cal H}_M$ as states in ${\cal H}_\gamma$ when the dissipation appears.
Notice that the prescription~(\ref{pres}) sends the vacuum in ${\cal H}_M$ to the vacuum in ${\cal H}_\gamma$,
and boils down to the identity map when $\gamma=0$.
The vacua define the measure in the $L^2$-scalar product, and 
(\ref{pres}) amounts essentially to make the substitution $p\rightarrow p_\gamma=\frac{p}{1-i2\gamma/\omega}$ 
for the momentum.

Therefore, according to our previous prescription, the state in ${\cal H}_\gamma$ that would describe the system at $t=0$
is:
\begin{equation}\label{statesO}
\psi^\gamma(q,p,0) = \frac{1}{\sqrt{n!}} (\bar b(q,p))^n \rho^\gamma_0(q,p).
\end{equation}
It is not an eigenstate of $H$ for the product $\star_\gamma$, but it is an element of ${\cal H}_\gamma$ and
therefore can be expanded  on the basis $\rho^\gamma_n$ given by (\ref{orthonor}) in ${\cal H}_\gamma$.
Using the following expression involving the Hermite polynomials:
$$
x^{n} =  \frac{n!}{2^{n}} \sum_{k=0}^{[n/2]} \frac{H_{n-2k}(x)}{k!(n-2k)!},\quad n\in\mathbb{N},
$$
where $[n/2]$ is the integer part of $n/2$, it is readily seen that the expansion of $\psi^\gamma(q,p,0)$ has the form:
\begin{equation}\label{psiexpand}
\psi^\gamma(q,p,0) = \sum_{k=0}^{[n/2]} \alpha^{(n)}_k \rho^\gamma_{n-2k}(q,p),
\end{equation}
We obtain after a simple computation
\begin{equation}\label{coeffa}
\alpha^{(n)}_k = \frac{\Gamma^k}{2^n k!}\sqrt{\frac{n!}{(n-2k)!}},
\quad k=0,1,\ldots,[n/2].
\end{equation}
Between $t=0$ and $t=\tau$, the state $\psi^\gamma$ evolves according to the dynamics defined by the product $\star_\gamma$
and we have:
\begin{equation}\label{psiexpandt}
\psi^\gamma(q,p,t) = \sum_{k=0}^{[n/2]} \alpha^{(n)}_k \exp\big(\frac{E_{n-2k}^\gamma t}{i\hbar}\big) 
\rho^\gamma_{n-2k}(q,p),\quad 0<t<\tau,
\end{equation}
where $E_{m}^\gamma = \hbar\omega(m + 1/2 + i\gamma/2\omega)$.

When $t=\tau$ the dissipation is turned off and the system will evolve as a harmonic oscillator in the Hilbert
space ${\cal H}_M$. We apply the inverse of the prescription~(\ref{pres}), i.e., 
\begin{equation}\label{presi}
\frac{1}{\sqrt{n!}} (\bar b(q,p))^n \rho^\gamma_0(q,p)\rightarrow  
\frac{1}{\sqrt{n!}} (\bar a(q,p))^n \rho_0(q,p)
\end{equation}
so that the state $\psi^\gamma$ at $t=\tau$ becomes an element $\psi$ of the Hilbert space ${\cal H}_M$. 
After applying the prescription~(\ref{presi}) and using the eigenstates~(\ref{orthonor}), we find
\begin{equation}\label{psiexpandtm}
\psi(q,p,\tau) = \sum_{k=0}^{[n/2]} \alpha^{(n)}_k \exp\big(\frac{E_{n-2k}^\gamma \tau}{i\hbar}\big) 
 \frac{\Gamma^{n/2 -k}}{\sqrt{(n-2k)!}} \
H_{n-2k}\big(\frac{1}{\Gamma^{1/2}}\ \bar a(q,p)\big)
\rho_0(q,p).
\end{equation}
Notice again that $\psi(q,p,\tau)$ is not an eigenstate of the Hamiltonian $H$ for the Moyal product.
It can be expressed in terms of the basis~(\ref{states}) of ${\cal H}_M$ with the help of the explicit
expression of the Hermite polynomials
$$
H_{n}(x) =  \sum_{j=0}^{[n/2]} (-1)^j \frac{n!}{(n- 2j)!j!}  (2x)^{n-2j}.
$$
A straightforward computation gives
\begin{equation}\label{psimoyaltau}
\psi(q,p,\tau) = \sum_{k=0}^{[n/2]} \beta^{(n)}_k(\tau) \rho_{n-2k}(q,p),
\end{equation}
where
\begin{equation}\label{beta}
\beta^{(n)}_k(\tau) = \frac{\Gamma^k}{2^{2k}k!}
\sqrt{\frac{n!}{(n-2k)!}} \exp(- i \omega(n+1/2 +i\gamma/2\omega )\tau) \big(\exp(2 i \omega \tau) - 1\big)^k,
\end{equation}
for $k=0,1,\ldots,[n/2]$. Notice that due to the nonunitary evolution during the damping regime, the state
$\psi(q,p,\tau)$ is no longer normalized. In order to maintain the probabilistic interpretation, 
$\psi$ should be normalized with respect to (\ref{scalar}). We denote by $\Psi$ the normalized state:
\begin{equation}\label{betan}
\Psi(q,p,\tau) = \frac{1}{N_n(\tau)}\sum_{k=0}^{[n/2]} \beta^{(n)}_k(\tau) \rho_{n-2k}(q,p)
\equiv \sum_{k=0}^{[n/2]} \hat\beta^{(n)}_k(\tau) \rho_{n-2k}(q,p),
\end{equation}
where $N_n(\tau)= \big(\sum_{k=0}^{[n/2]} |\beta^{(n)}_k(\tau)|^2\big)^{1/2}$. For $t>\tau$, the evolution
of the state $\Psi$ is given as usual by:
\begin{equation}\label{psit}
\Psi(q,p,t) = \sum_{k=0}^{[n/2]} \hat\beta^{(n)}_k(\tau) \exp(- i\omega(n-2k+1/2 )(t-\tau))\rho_{n-2k}(q,p),\quad t>\tau .
\end{equation}

In view of (\ref{betan}), we notice that for generic values of $\tau$, the state of energy $\hbar\omega(n+1/2)$ we started with can decay
to states of lower energy level $n-2k$ of same parity as $n$ with transition probability $|\hat\beta^{(n)}_k(\tau)|^2$. 
Hence the state of energy $E_n$ spreads over lower energy states and the expectation value for the mechanical
energy $H$ would be less than $E_n$ after the system has experienced dissipation.
However, for specific values of $\tau$, i.e. when $\omega \tau =
\pi + 2m\pi$, $m\in\mathbb{N}$, we have $\hat\beta^{(n)}_k(\tau) = \delta_{0k}\exp(- i\omega(n+1/2 )\tau)$ and the system
is still in the state $\rho_n$ and its behavior seems not be affected by the dissipation.

\section{Conluding remarks}
Taking as a model the quantum description of the harmonic oscillator in deformation 
quantization, we have obtained a quantization for the damped harmonic oscillator in terms of a 
new star product $\star_\gamma$. This product can be viewed as a deformation in two parameters, 
$\hbar$ and $\gamma$, that allows  to recover at any point the classical case when 
$\hbar \rightarrow 0$ and the usual quantized harmonic oscillator when $\gamma \rightarrow 0$. 

The simple toy-model considered in Sect.~4 can be replaced by a more physical one. Instead
of switching on/off abruptly the dissipation, a more physical situation would correspond to switching on/off
the dissipation by letting $\gamma$ depend on time. $\gamma(t)$ would then be a smooth
positive function of $t$ with support in some time interval $[\tau_0,\tau_1]$. The star-product
$\star_{\gamma(t)}$ would also depend on time and  is defined exactly as in (\ref{dhoseven}) as the time-dependence
of $\gamma$ does not alter the associativity. 
For a time $s$ not in the support of  $\gamma(t)$, $\star_{\gamma(s)}$ is simply the Moyal product and the whole
system can be quantized with a unique time-dependent star-product.

It is expected that the results obtained here will give more  physical 
applications of the deformation quantization formalism and motivate the study of more 
complex systems.


\section*{Acknowledgments}
We would like to thank Bogdan Mielnik and Maciej Przanowski for critical remarks on the manuscript
and useful discussions we had with them when the authors were visiting the CINVESTAV in Mexico City.
This paper was partially supported by the CONACYT grant 41993-F. One of us (F.T.) was partially supported 
by the CONACYT postdoctoral scholarship program and a postdoctoral fellowship from the R\'egion de Bourgogne.
F.T. wants to thank the Institut de Math\'ematiques de Bourgogne (France) for their warm 
hospitality.


\end{document}